\documentclass{article}
\usepackage{kr}

\usepackage{times}
\usepackage{soul}
\usepackage{url}
\usepackage[hidelinks]{hyperref}
\usepackage[utf8]{inputenc}
\usepackage[small]{caption}
\usepackage{graphicx}
\usepackage{amsmath}
\usepackage{amsthm}
\usepackage{booktabs}
\usepackage{amssymb}
\usepackage{enumitem}
\usepackage{xcolor}
\usepackage{xspace}
\usepackage{tikz}
\usetikzlibrary{arrows.meta, calc, positioning, shapes.geometric}
\usepackage[capitalize]{cleveref}
\usepackage[bibliography=common]{apxproof}

\newtheoremrep{theorem}{Theorem}
\newtheoremrep{lemma}[theorem]{Lemma}
\newtheoremrep{proposition}[theorem]{Proposition}
\newtheoremrep{corollary}[theorem]{Corollary}
\newtheorem{definition}[theorem]{Definition}

\definecolor{lilac}{rgb}{0.8, 0.4, 0.7}
\definecolor{fox}{rgb}{0.9, 0.5, 0.3}

\newcommand{\FormatRuleClass}[1]{\ensuremath{\mathtt{#1}}\xspace}
\newcommand{\bts}{\FormatRuleClass{bts}}
\newcommand{\gbts}{\FormatRuleClass{gbts}}
\newcommand{\fc}{\FormatRuleClass{fc}}
\newcommand{\bdd}{\FormatRuleClass{bdd}}
\newcommand{\fes}{\FormatRuleClass{fes}}
\newcommand{\fus}{\FormatRuleClass{fus}}

\newcommand{\adom}{\mathsf{adom}}
\newcommand{\body}{\mathsf{body}}
\newcommand{\head}{\mathsf{head}}
\newcommand{\fr}{\mathsf{fr}}
\newcommand{\tw}{\mathsf{tw}}

\newcommand{\pa}{\pi}
\newcommand{\apply}{\mathsf{apply}}
\newcommand{\chase}[1]{\mathsf{Ch}(#1)}
\newcommand{\chasedr}{\chase{\db,\rs}}
\newcommand{\step}[2]{\mathsf{Ch}_{#1}(#2)}
\newcommand{\stepdr}[1]{\step{#1}{\db,\rs}}

\newcommand{\chaseo}{\chase{\db_0,\rs}}
\newcommand{\stepo}[1]{\step{#1}{\db_0,\rs}}
\newcommand{\nextStep}[1]{\mathsf{Next}_{\rs}(#1)}
\newcommand{\pred}[1]{\mathrm{#1}}
\definecolor{colE}{HTML}{268BD2}
\definecolor{colSame}{HTML}{C27C4E}
\definecolor{colA}{HTML}{B5466F}
\newcommand{\E}{\textcolor{colE}{\mathsf{E}}}
\newcommand{\A}{\textcolor{colA}{\mathsf{A}}}
\newcommand{\same}{\textcolor{colSame}{\mathsf{Same}}}

\newcommand{\nats}{\mathbb{N}}

\newcommand{\querylang}{\mathfrak{Q}}
\newcommand{\ontlang}{\mathfrak{O}}

\newcommand{\pair}[1]{\langle#1\rangle}
\newcommand{\set}[1]{\{#1\}}

\newcommand{\db}{\mathcal{D}}
\newcommand{\rs}{\mathcal{R}}

\title{Bounded Treewidth Rulesets Are Not Finitely Controllable}

\author{%
Piotr Ostropolski-Nalewaja$^{1}$\and
Micha\"el Thomazo$^2$\\
\affiliations
$^1$University of Wroc{\l}aw, Poland\\
$^2$Inria, DI ENS, ENS, CNRS, PSL University, Paris, France\\
\emails
postropolski@cs.uni.wroc.pl,
michael.thomazo@inria.fr
}

\begin{document}

\maketitle

\begin{abstract}
    An ontology language $\ontlang$ is \emph{finitely controllable} with respect to a query language $\querylang$ if the finite and unrestricted variants of the query entailment problem for $\pair{\ontlang, \querylang}$ coincide. In this paper, we study finite controllability (\fc) in its classical sense for existential rules, where $\ontlang$ is a class of existential rules and $\querylang$ is the class of all Boolean conjunctive queries (BCQs).

    A natural question in this line of work is whether the class of bounded treewidth sets (\bts) is finitely controllable with respect to BCQs. We answer this question negatively: we exhibit a simple set of rules over a binary signature and prove that it is \bts but not \fc. Moreover, using the same set of rules, we show that two subclasses of \bts, namely \gbts and pieceful, are not \fc either.
\end{abstract}

\section{Introduction}\label{sec:intro}

This paper concerns two central topics in knowledge representation:
\emph{existential rules} and \emph{finite controllability}.

\subsubsection{Existential Rules}
Existential rules, also called \emph{tuple-generating dependencies} (TGDs),
$\forall\exists$-rules, or Datalog$^{\pm}$ rules, are first-order sentences of
the form $\forall\bar x\,\forall\bar y\,
(\alpha(\bar x,\bar y)\to\exists\bar z\,\beta(\bar y,\bar z))$,
where $\alpha$ and $\beta$ are conjunctions of relational
atoms~\cite{Abiteboul1995,Baget2011,Cali2009}.

When combined with a database, such rules express background knowledge,
potentially requiring the existence of elements not named in the data.
The fundamental reasoning task for such a combination is
\emph{ontology-mediated query answering} (OMQA): given a finite database $\db$,
a ruleset $\rs$, and a Boolean conjunctive query $q$, decide whether
$\db,\rs\models q$.
Without restrictions on the rules, this problem is undecidable.

A particularly useful tool is the \emph{chase} algorithm. Starting from $\db$,
it repeatedly applies rules, adding atoms and fresh elements witnessing
existential conclusions. Crucially, the chase is a \emph{universal model}:
it admits a homomorphism into every model of $\pair{\db,\rs}$, so CQ entailment
reduces to evaluating the query on this single structure~\cite{Deutsch2008,DBLP:journals/jacm/BeeriV84,DBLP:journals/tods/MaierMS79}.
Its clear procedural definition and tight connection to query answering are
reflected in a wide array of practical reasoning tools, such as Graal,
InteGraal, VLog, Nemo, and Vadalog~\cite{Graal2015,InteGraal2023,VLog2018,Nemo2024,Vadalog2018}.

\subsubsection{Knowledge Representation in the Finite}
Given a database $\db$ and a ruleset $\rs$, query entailment admits two
settings: the usual, \emph{unrestricted} one, which considers all models of
$\pair{\db, \rs}$, and the \emph{finite} one, which considers only finite structures.
Interestingly, while the finite setting is arguably the natural choice for
computer science, the unrestricted setting is standard in knowledge
representation. There are, however, good reasons for this.
In general, the unrestricted setting offers a richer and better-explored
mathematical toolkit---classical examples are the L\"owenheim--Skolem and
Compactness theorems, which do not hold when restricted to finite
structures~\cite{Libkin2004}.
From the perspective of existential rules, and of this paper in particular,
the finite setting also breaks the elegant connection between query
entailment and evaluating the query on the chase.
Insisting on finite models from the outset would sacrifice this advantage.
Therefore, a more fruitful approach is to study whether restricting to
finite models changes the answers at all.

A ruleset $\rs$ is \emph{finitely controllable} (\fc) if, for every finite
database $\db$ and Boolean CQ $q$,
\[
 \db,\rs \models q
 \quad\Longleftrightarrow\quad
 \db,\rs \models_{\mathrm{fin}} q.
\]
In other words, finite controllability allows us to assume the natural
finite setting while retaining the soundness and completeness of
unrestricted reasoning tools---combining the best of the finite and
unrestricted worlds.
Moreover, although this is not the focus of the present paper,
\fc~is also an important decidability criterion for CQ
entailment~\cite{Feller2023}.

\subsubsection{Classes of Existential Rules}
Before discussing the intersection of existential rules and finite
controllability, we need to introduce some important classes of existential
rules, arising from research on OMQA.

In general, OMQA is undecidable for existential rules; thus, a major line
of research seeks classes of existential rules for which OMQA is decidable.
Broadly, there are two types of such classes:
concrete, syntactically defined classes that come with decidable membership
tests, and broader, semantically defined classes that capture underlying
\emph{reasons} for decidability---typically at the price of undecidable
membership~\cite{Baget2011}.
Two important such reasons are \emph{first-order rewritability} and
\emph{structural restrictions on universal models}.

\medskip
\noindent
\textbf{Finite Unification Sets.}
The first reason is captured by \emph{finite unification sets} (\fus), also known
through the \emph{bounded derivation depth} (\bdd) property.
For \fus~rulesets, each CQ can be rewritten into a union of CQs and then
evaluated directly over the input database, completely bypassing the ruleset.
Linear and sticky rules are prominent syntactic examples of this
approach~\cite{Cali2009,Baget2011,Gogacz2013}.

\medskip
\noindent
\textbf{Bounded Treewidth Sets.}
The second approach, closely related to this paper, constrains the shape
of a universal model.
Here, the most basic semantically defined class is that of
\emph{finite expansion sets} (\fes), which admit finite universal models.
The next step, going beyond finiteness, is the class of
\emph{bounded-treewidth sets} (\bts), whose rulesets admit universal models
of finite treewidth for every database~\cite{Baget2023PODS}.
This class also supports decidable CQ entailment even when such a
model is infinite~\cite{Baget2011,Feller2023}.
While \bts is the most recognized class in this line, the idea of requiring
a universal model of finite width can be generalized further to
\emph{finite-cliquewidth sets} and \emph{finite-partitionwidth
sets}~\cite{Feller2023,Feller2025}.

Important classes falling within \bts include the
\emph{guarded} family---\emph{guarded}, \emph{frontier-guarded}, and
\emph{weakly guarded} rules---which constrain rule shapes; as well as
\emph{greedy bounded-treewidth sets} (\gbts) and their subclass of
\emph{pieceful} rulesets, which constrain chase
derivations~\cite{Baget2011GBTS,Buron2021}.

\subsubsection{Existential Rules and Finite Controllability}
Research on finite controllability of existential rules is well advanced
but far from complete. In general, \fc is known to hold for many
syntactically defined classes on both the \fus and \bts fronts, but the
corresponding questions for these two semantic classes have remained open.
Below we discuss the current research landscape.

\medskip
\noindent
\textbf{FC and Finite Unification Sets.}
On the rewriting side, whether every \fus/\bdd ruleset is \fc is the
long-standing \emph{\bdd/\fc conjecture}, still open despite considerable
effort~\cite{Gogacz2013,Larroque2025}.
The major syntactic subclass of \fus known as \emph{sticky} rules
is \fc~\cite{Gogacz2017}.

\medskip
\noindent
\textbf{FC and Bounded Treewidth Sets.}
On the structural side, \fc follows immediately for the semantic class
\fes, but no analogous general result was known for \bts.
As on the \fus side, however, there is a range of results for syntactically
defined classes. Inclusion dependencies~\cite{Rosati2011},
guarded and weakly guarded rules~\cite{Barany2014}, and frontier-guarded
rules~\cite{Barcelo2020} are known to be \fc.

However, unlike for \fus, there is a considerable gap between full \bts
and the classes known to be \fc. In particular, neither the pieceful
class nor the \gbts class was known to be \fc.
It is therefore natural to study where the boundary of finite
controllability lies within the structural landscape of \bts and beyond.

\emph{We believe that these two finite-controllability questions, concerning
\fus and \bts, are among the most interesting in the study of existential
rules.} In this paper, we settle one of them: whether \bts implies \fc.

\subsubsection{Our Contribution}
In this work, we give a simple counterexample to the claim that \bts implies \fc.
Specifically, we exhibit a pieceful ruleset (and thus a \bts ruleset)
that is not \fc. Consequently, we close the picture regarding finite controllabilty and the structural front.

\section{Preliminaries}\label{sec:prelims}

\paragraph{Atoms and instances.}
We consider pairwise disjoint, countably infinite sets of \emph{constants}, \emph{variables} and \emph{nulls}, and a set of \emph{predicates}, each with an arity.
A \emph{signature} is a finite set of predicates, and it is \emph{binary} if all its predicates have arity at most~$2$.
A set of atoms, a rule, a ruleset or a query is \emph{over} a signature if every predicate occurring in it belongs to this signature.
A \emph{term} is a constant, a variable or a null.
An \emph{atom} is an expression $\pred{P}(t_1,\dots,t_n)$ where $\pred{P}$ is an $n$-ary predicate and $t_1,\dots,t_n$ are terms.
The \emph{active domain} $\adom(S)$ of a set of atoms $S$ is the set of terms occurring in $S$.
An \emph{instance} is a possibly infinite set of atoms over constants and nulls, and a \emph{database} is a finite set of atoms over constants.
A \emph{homomorphism} from a set of atoms $S$ to a set of atoms $S'$ is a map $h:\adom(S)\to\adom(S')$ that is the identity on constants and such that $h(S)\subseteq S'$.
For a natural number $n$, we write $[n]$ for the set $\set{0, \ldots, n}$.

\paragraph{Queries.}
A \emph{(Boolean) conjunctive query} (CQ) is a formula $q=\exists\vec{x}\;\varphi(\vec{x})$, where $\varphi$ is a conjunction of atoms over variables.
An instance $I$ \emph{satisfies} $q$, written $I\models q$, if there is a homomorphism from the atoms of $\varphi$ to $I$.
We write $\db,\rs\models q$ if every model of $\db$ and $\rs$ satisfies $q$.
A query $q$ is \emph{homomorphism closed} if for any two instances $I$ and $I'$ such that $I$ homomorphically maps to $I'$ it holds that $I \models q \Rightarrow I' \models q$.
It is folklore that BCQs are homomorphism closed.

\subsection{Existential Rules}
An \emph{(existential) rule} $\rho$ is a formula
\[
\forall\vec{x}\,\forall\vec{y}\;\big(B(\vec{x},\vec{y})\to\exists\vec{z}\;H(\vec{y},\vec{z})\big),
\]
where $B$ and $H$ are non-empty conjunctions of atoms over variables.
We write $\body(\rho)$ and $\head(\rho)$ for the sets of atoms of $B$ and $H$.
The \emph{frontier} of $\rho$ is the set $\fr(\rho)=\vec{y}$ of variables shared by its body and its head, and $\vec{z}$ are its \emph{existential variables}.
A rule is \emph{Datalog} if it has no existential variable.
We omit universal quantifiers when writing rules.
An instance $I$ \emph{satisfies} $\rho$ if every homomorphism from $\body(\rho)$ to $I$ extends to a homomorphism from $\body(\rho)\cup\head(\rho)$ to $I$.
A \emph{model} of a database $\db$ and a ruleset $\rs$ is an instance that contains $\db$ and satisfies every rule of $\rs$.

\paragraph{Triggers.}
A \emph{trigger} for an instance $I$ is a pair $(\rho,h)$ where $\rho$ is a rule and $h$ is a homomorphism from $\body(\rho)$ to $I$. An application of a trigger $\pi = (\rho, h)$ for an instance $I$, denoted $\apply(\pi, I)$, is a set of atoms obtained from the $\head(\rho)$ by replacing each universally quantified variable $x$ with $h(x)$, and replacing each existentially quantified variable with a fresh null. The trigger $\pi$ is \emph{inactive} if there is an extension $h'$ of $h$ over the existentially quantified variables of $\rho$ such that $h'$ is a homomorphism from $\head(\rho) \cup \body(\rho)$ to $I$. A trigger is active if it is not inactive.

\paragraph{Derivations.}
A \emph{derivation} from a database $\db$ with a ruleset $\rs$ is a finite or infinite sequence $\db=I_0\subseteq I_1\subseteq\cdots$ such that, for every step $i\geq 1$, there is a trigger $\pi$ for $I_{i-1}$ such that $I_{i} = I_{i - 1} \cup \apply(\pi, I_{i-1})$. An instance is \emph{derived} from $\db$ with $\rs$ if it is an element of some derivation from $\db$ with $\rs$.
A derivation is \emph{fair} if, for every $i$ and every trigger $\pi$ for $I_i$ there is $j\geq i$ such that $I_{j + 1} = I_{j} \cup \apply(\pi, I_{j})$. \emph{We specifically use derivations for definition of \gbts and pieceful classes (\cref{def:pieceful}).}

\paragraph{The Chase}
From the perspective of the main result of this paper, it is easier to introduce the chase algorithm. Take a database $\db$ and a ruleset $\rs$. For an instance $I$, we denote by $\nextStep{I}$ the instance $I \cup \bigcup_{\pi \in A} \apply(\pi, I)$, where $A$ is the set of active triggers for the rules of $\rs$ in $I$. Let $\stepdr{0} = \db$ and $\stepdr{i+1} = \nextStep{\stepdr{i}}$ for each natural $i$; we call $\stepdr{i}$ the \emph{$i$-th step of the chase}. The \emph{chase} $\chasedr$ is the instance $\bigcup_{i = 0}^{\infty}\stepdr{i}$.

\paragraph{Universal models.}
A \emph{universal model} of $\db$ and $\rs$ is a model of $\db$ and $\rs$ that maps homomorphically into every model of $\db$ and $\rs$. It is well known that the chase is a \emph{universal model} of $\pair{\db, \rs}$~\cite{Deutsch2008,DBLP:journals/jacm/BeeriV84,DBLP:journals/tods/MaierMS79}
Hence, given any homomorphism closed query $q$, $\db,\rs\models q$ if and only if the chase satisfies $q$.

\subsection{Treewidth and Related Classes of Existential Rules.}
The rulesets we study are defined by requiring derived instances to be tree-like, a notion formalised by treewidth.
The \emph{Gaifman graph} of an instance $I$ has vertex set $\adom(I)$, and an edge between any two distinct terms that occur together in an atom of $I$.
A \emph{tree decomposition} of a graph $G$ is a pair $(T,(B_n)_{n\in T})$ where $T$ is a possibly infinite tree and each \emph{bag} $B_n$ is a set of vertices of $G$, such that every edge of $G$ is contained in some bag, and for every vertex $v$ of $G$ the nodes $n$ with $v\in B_n$ form a non-empty connected subtree of $T$.
Its \emph{width} is $\sup_{n\in T}|B_n|-1$.
The \emph{treewidth} $\tw(I)$ of $I$ is the minimal width of a tree decomposition of its Gaifman graph.

We consider the following properties of rulesets.
Bounded treewidth sets~\cite{Baget2011} are rulesets whose derived instances are tree-like.

\begin{definition}\label{def:bts}
A ruleset $\rs$ is a \emph{bounded treewidth set} (\bts) if for every database $\db$ there is $b\in\nats$ such that $\db$ and $\rs$ admit a universal model of treewidth at most $b$.
\end{definition}

Greedy~\cite{Baget2011GBTS} and pieceful~\cite{Buron2021} rulesets are subclasses of \bts defined by restricting where rule applications map their frontier.

\begin{definition}\label{def:pieceful}
Let $\db=I_0\subseteq I_1\subseteq\cdots$ be a derivation from a database $\db$.
For every step $i\geq1$, let $(\rho_i,h_i)$ be the trigger applied at step $i$, and let $A_i=\apply((\rho_i,h_i),I_{i-1})$ be the set of atoms it produces, so that $I_i=I_{i-1}\cup A_i$.
It is \emph{greedy} if, for every step $i$, there is $1 \leq j<i$ such that $h_i(\fr(\rho_i))\subseteq\adom(\db)\cup\adom(A_j)$.
It is \emph{pieceful} if, for every step $i$:
\begin{itemize}
\item either $h_i(\fr(\rho_i))\subseteq\adom(\db)$;
\item or $h_i(\fr(\rho_i))\subseteq\adom(A_j)$ for some $j<i$.
\end{itemize}
A ruleset is \emph{\gbts} (resp.\ \emph{pieceful}) if every derivation from every database with this ruleset is greedy (resp.\ pieceful).
\end{definition}

Every pieceful derivation is greedy, so pieceful rulesets are \gbts.
Moreover, \gbts rulesets are \bts~\cite{Baget2011GBTS}.

Finally, finite controllability~\cite{DBLP:journals/jcss/JohnsonK84,Rosati2011} states that restricting to finite models does not change CQ entailment.

\begin{definition}\label{def:fc}
A ruleset $\rs$ is \emph{finitely controllable} (\fc) if for every database $\db$ and every CQ $q$ such that $\db,\rs\not\models q$, there is a finite model $M$ of $\db$ and $\rs$ such that $M\not\models q$.
\end{definition}

\section{The Counterexample}\label{sec:counterexample}

\begin{theorem}\label{thm:main}
There is a \bts ruleset over a binary signature that is not \fc.
\end{theorem}

Let $\rs$ consist of the following rules over the binary predicates $\E$ and $\same$:
\begin{align}
\E(x,y) &\to \exists z\; \E(y,z) \tag{$\rho_1$}\label{r1}\\
\E(x,y) &\to \same(y,y) \tag{$\rho_2$}\label{r2}\\
\same(y,y')\wedge \E(x,y)\wedge \E(x',y') &\to \same(x,x') \tag{$\rho_3$}\label{r3}
\end{align}

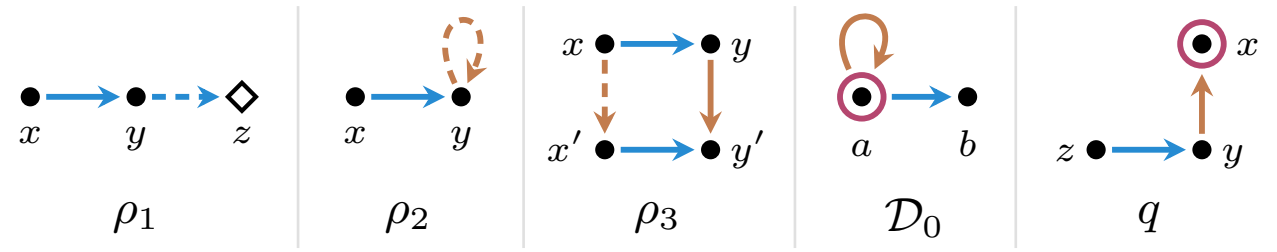
\begin{figure}[t]
\centering
\begin{tikzpicture}[
    every node/.style={inner sep=0pt},
    uvar/.style={circle, fill=black, minimum size=3.6pt},
    evar/.style={diamond, draw=black, fill=white, line width=0.8pt, minimum size=5.6pt},
    lbl/.style={font=\scriptsize, inner sep=1pt, text height=1.2ex, text depth=0.25ex},
    arr/.style={line width=1.1pt, -{Stealth[length=4.2pt, width=4.2pt]}, shorten >=1pt, shorten <=1pt},
    E/.style={arr, draw=colE},
    S/.style={arr, draw=colSame},
    head/.style={dash pattern=on 2.6pt off 1.8pt},
    Aring/.style={circle, draw=colA, line width=1.1pt, minimum size=8.2pt},
    pname/.style={font=\small, inner sep=0pt},
    sep/.style={draw=black!12, line width=0.5pt},
]
\def\L{0.7}    
\def\H{0.35}   
\def\Y{-0.8}   
\def\sameloop#1#2#3{\draw[S, #1, shorten <=#3, shorten >=#3] (#2.center) .. controls +(115:0.7) and +(65:0.7) .. (#2.center);}

\begin{scope}[local bounding box=p1]
    \node[uvar] (x) at (0,0) {};
    \node[uvar] (y) at (\L,0) {};
    \node[evar] (z) at (2*\L,0) {};
    \draw[E] (x) -- (y);
    \draw[E, head] (y) -- (z);
    \node[lbl, below=2pt of x.center] {$x$};
    \node[lbl, below=2pt of y.center] {$y$};
    \node[lbl, below=2pt of z.center] {$z$};
    \node[pname] at (\L,\Y) {$\rho_1$};
\end{scope}

\begin{scope}[xshift=2.15cm, local bounding box=p2]
    \node[uvar] (x) at (0,0) {};
    \node[uvar] (y) at (\L,0) {};
    \draw[E] (x) -- (y);
    \sameloop{head}{y}{2.8pt}
    \node[lbl, below=2pt of x.center] {$x$};
    \node[lbl, below=2pt of y.center] {$y$};
    \node[pname] at (\L/2,\Y) {$\rho_2$};
\end{scope}

\begin{scope}[xshift=3.8cm, local bounding box=p3]
    \node[uvar] (x)  at (0,\H)   {};
    \node[uvar] (y)  at (\L,\H)  {};
    \node[uvar] (xp) at (0,-\H)  {};
    \node[uvar] (yp) at (\L,-\H) {};
    \draw[E] (x) -- (y);
    \draw[E] (xp) -- (yp);
    \draw[S] (y) -- (yp);
    \draw[S, head] (x) -- (xp);
    \node[lbl, left=0.5pt of x]   {$x$};
    \node[lbl, left=0.5pt of xp]  {$x'$};
    \node[lbl, right=0.5pt of y]  {$y$};
    \node[lbl, right=0.5pt of yp] {$y'$};
    \node[pname] at (\L/2,\Y) {$\rho_3$};
\end{scope}

\begin{scope}[xshift=5.5cm, local bounding box=p4]
    \node[uvar] (a) at (0,0) {};
    \node[uvar] (b) at (\L,0) {};
    \node[Aring] (ar) at (a) {};
    \draw[E] (ar) -- (b);
    \sameloop{}{a}{5.7pt}
    \node[lbl, below=4.5pt of b.center -| ar] {$a$};
    \node[lbl, below=4.5pt of b.center]  {$b$};
    \node[pname] at (\L/2,\Y) {$\db_0$};
\end{scope}

\begin{scope}[xshift=7.05cm, local bounding box=p5]
    \node[uvar] (z) at (0,-\H) {};
    \node[uvar] (y) at (\L,-\H) {};
    \node[uvar] (x) at (\L,\H) {};
    \node[Aring] (xr) at (x) {};
    \draw[E] (z) -- (y);
    \draw[S] (y) -- (xr);
    \node[lbl, left=0.5pt of z]   {$z$};
    \node[lbl, right=0.5pt of y]  {$y$};
    \node[lbl, right=0.5pt of xr] {$x$};
    \node[pname] at (\L/2,\Y) {$q$};
\end{scope}

\foreach \l/\r in {p1/p2, p2/p3, p3/p4, p4/p5}
    \path ($(\l.east)!0.5!(\r.west)$) coordinate (m) [sep] (m |- 0,-1.0) edge (m |- 0,0.6);
\end{tikzpicture}
\caption{The rules of $\rs$, the database $\db_0$, and the query $q$. Blue arrows denote $\E$-atoms, orange arrows $\same$-atoms, and raspberry rings $\A$-atoms. Dashed arrows are head atoms; all other atoms are solid. Diamonds denote existentially quantified variables of rule heads; bullets denote all other variables and constants.}
\label{fig:rules}
\end{figure}

The intuition behind the above rules is simple. The first rule \emph{builds an infinite $\E$-path} in the chase. The second rule ensures that every term\footnote{To be precise, every term that has an incoming $\E$-edge.} \emph{is on the same level} as itself, and the third rule propagates this information backwards: two terms are on the same level if their successors are.

We now define the database and the query that, together with $\rs$, witness \cref{thm:main}.

Let $\db_0=\{\A(a),\same(a,a),\E(a,b)\}$, where $\A$ is a unary predicate, and let
\[
q=\exists x\,\exists y\,\exists z\;\big(\A(x)\wedge\same(y,x)\wedge \E(z,y)\big).
\]

Note that in light of the above intuition, the query $q$ checks if there exists a term with an incoming $\E$-edge that is on the same level as some term satisfying the predicate $\A$.

We now describe the shape of the chase of $\db_0$ and $\rs$.

\begin{lemma}\label{obs:chase-def}
    There is an infinite sequence $t_0, t_1, t_2, \ldots$ of pairwise distinct terms, with $t_0 = a$, $t_1 = b$, and $t_j$ a null for every $j \geq 2$, such that for every natural number $i$, the instance $\stepo{i}$ is equal to:\footnote{We use $\db_0$ in the formula only for notational convenience; it can be freely replaced by $\set{\A(a)}$.}
    \[
    \db_0 \cup \set{\E(t_j, t_{j + 1}) \mid j \in [i]} \cup \set{\same(t_j, t_j) \mid j \in [i]}.
    \]
\end{lemma}
\begin{proof}[Proof sketch]
    By induction on $i$; the base case $i = 0$ is immediate. For the induction step, observe that $\stepo{i}$ has exactly two active triggers: one for $\rho_1$ and one for $\rho_2$, both mapping the body atom to the last edge $\E(t_i, t_{i+1})$. All other triggers for $\rho_1, \rho_2$ and $\rho_3$ are already satisfied. Applying the two active triggers yields $\stepo{i+1}$ of the required form. The full proof is in \cref{apx:chase-def}.
\end{proof}

It should come as no surprise that the query $q$ does not hold in $\chaseo$.

\begin{lemma}
    $\chaseo \not\models q$.
\end{lemma}
\begin{proof}
    By \cref{obs:chase-def}, the only $\same$-atoms of $\chaseo$ are $\same(t_j, t_j)$, and $a = t_0$ has no incoming $\E$-edge. Suppose $h$ is a homomorphism from $q$ to $\chaseo$. As $a$ is the only term satisfying $\A$, we have $h(x) = a$, hence $h(y) = a$, and so $\E(h(z), a) \in \chaseo$, a contradiction.
\end{proof}

As $\chaseo$ is a model of $\pair{\db_0, \rs}$, we know that $q$ is not entailed in general. We now show that, in contrast, $q$ holds in every finite model of $\pair{\db_0, \rs}$.

\paragraph{Finite Entailment}

Consider any finite model $M$ of $\pair{\db_0, \rs}$. As $\chaseo$ is a universal model, there exists a homomorphism $h$ from $\chaseo$ to $M$. For each natural $i$ denote $h(t_i)$ as $m_i$.

Note the following trivial lemma:
\begin{lemma}\label{lem:easy-step}
    For all naturals $i, j > 0$, if $M \models \same(m_i, m_j)$, then $M \models \same(m_{i - 1}, m_{j - 1})$.
\end{lemma}
\begin{proof}
    As $h$ is a homomorphism and $\E(t_{i-1}, t_i), \E(t_{j-1}, t_j) \in \chaseo$ by \cref{obs:chase-def}, we have $M \models \E(m_{i-1}, m_i) \wedge \E(m_{j-1}, m_j)$. Together with $\same(m_i, m_j)$, this matches the body of $\rho_3$, so $M \models \same(m_{i-1}, m_{j-1})$ as $M$ satisfies $\rho_3$.
\end{proof}

As $M$ is finite, there exist naturals $c < c'$ such that $m_c = m_{c'}$. Note that $M \models \same(m_{c'}, m_c)$, as $\same(t_c, t_c) \in \chaseo$ is mapped by $h$ to $\same(m_c, m_c) = \same(m_{c'}, m_c)$.

Recall that $c < c'$ and let $d = c' - c > 0$. We show that $M \models \same(m_d, m_0)$. If $c > 0$, then applying \cref{lem:easy-step} $c$ times yields $M \models \same(m_{c'-c}, m_{c-c}) = \same(m_d, m_0)$. Assume now $c = 0$, so $d = c'$, and thus $M \models \same(m_d, m_0)$. Since $h(a) = a$, we have $m_0 = a$, and since $\E(t_{d-1}, t_d) \in \chaseo$, we have $M \models \E(m_{d-1}, m_d)$. Hence $M \models \A(a) \wedge \same(m_d, a) \wedge \E(m_{d-1}, m_d)$, and thus $M \models q$.

Hence $\pair{\db_0, \rs} \not\models q$, yet $q$ holds in every finite model of $\pair{\db_0, \rs}$, which by definition gives the following.

\begin{corollary}\label{cor:sigma-not-fc}
$\rs$ is not \fc.
\end{corollary}

\section{Why \texorpdfstring{$\rs$}{R} Is Pieceful}\label{sec:pieceful}

Rules \eqref{r1} and \eqref{r2} have a single frontier variable, so the only threat to piecefulness is \eqref{r3}, whose frontier is mapped to $\E$-predecessors of two $\same$-related terms.
These terms may be distinct, as $m_d$ and $m_0$ in the finite-entailment argument of \cref{sec:counterexample}.
We show that in a derivation, this only happens on constants: $\same$ relates a null only to itself (\cref{lem:invariants}).
Fix a derivation $\db=I_0\subseteq I_1\subseteq\cdots$ from a database $\db$ with $\rs$, with the notation of \cref{sec:prelims}.

\begin{lemma}\label{lem:invariants}
For every $k$:
\begin{enumerate}[label=(I\arabic*)]
\item\label{inv:E} Every $\E$-atom of $I_k$ whose second argument is a constant belongs to $\db$. Every null $n$ of $I_k$ is the second argument of exactly one $\E$-atom of $I_k$, written $\E(\pa(n),n)$.
\item\label{inv:same} Every $\same$-atom of $I_k$ is either a loop $\same(t,t)$, or $\same(c,c')$ with $c,c'\in\adom(\db)$.
\item\label{inv:r3} If step $k$ applies \eqref{r3}, then either $h_k(x),h_k(x')\in\adom(\db)$, or $h_k(x)=h_k(x')$.
\end{enumerate}
\end{lemma}

\begin{proof}
\ref{inv:E}: Only \eqref{r1} produces $\E$-atoms or creates nulls, and each step applying it produces a single atom $\E(t,n)$, where $n$ is the null it creates.
Since $\db$ contains no null, the $\E$-atoms of $I_k$ with a constant as second argument are those of $\db$, and the only one with a null $n$ as second argument is produced by the step creating $n$.

\ref{inv:same} and \ref{inv:r3}: We proceed by simultaneous induction on $k$.
For $k=0$, \ref{inv:same} holds since $I_0=\db$, and \ref{inv:r3} is void.
Let $k\geq1$.
If step $k$ applies \eqref{r1} or \eqref{r2}, \ref{inv:r3} is void and \ref{inv:same} is preserved: \eqref{r1} produces no $\same$-atom, and \eqref{r2} only produces loops.
Assume now that step $k$ applies \eqref{r3}.
The atom $\same(h_k(y),h_k(y'))$ belongs to $I_{k-1}$, so by \ref{inv:same} for $I_{k-1}$, one of the following holds:
\begin{itemize}
\item \emph{$h_k(y)$ and $h_k(y')$ are constants} (this includes loops on constants).
By \ref{inv:E}, the body atoms $\E(h_k(x),h_k(y))$ and $\E(h_k(x'),h_k(y'))$ belong to $\db$, so $h_k(x),h_k(x')\in\adom(\db)$.
\item \emph{$h_k(y)=h_k(y')=n$ for a null $n$.}
By \ref{inv:E}, the body atoms $\E(h_k(x),n)$ and $\E(h_k(x'),n)$ are the same atom, so $h_k(x)=h_k(x')=\pa(n)$.
\end{itemize}
This proves \ref{inv:r3} for step $k$.
The produced atom $\same(h_k(x),h_k(x'))$ is over $\adom(\db)$ in the first case and a loop in the second, so \ref{inv:same} holds for $I_k$.
\end{proof}

\begin{proposition}\label{prop:sigma-pieceful}
$\rs$ is pieceful, hence \gbts, hence \bts.
\end{proposition}

\begin{proof}
By the discussion following \cref{def:pieceful}, it suffices to show that this derivation is pieceful.
Fix a step $i\geq1$.
It suffices to show that $h_i(\fr(\rho_i))$ is either contained in $\adom(\db)$ or a singleton $\{t\}$. Rules \eqref{r1} and \eqref{r2} have a single frontier variable, so $h_i(\fr(\rho_i))$ is a singleton.
Rule \eqref{r3} has frontier $\{x,x'\}$, and \ref{inv:r3} of \cref{lem:invariants} gives exactly the required dichotomy.
\end{proof}

\begin{proof}[Proof of \cref{thm:main}]
By \cref{prop:sigma-pieceful,cor:sigma-not-fc}, the ruleset $\rs$ over the binary signature $\{\E,\same\}$ is \bts and not \fc.
\end{proof}


\section{Conclusion}\label{sec:conclusion}

We have shown that bounded treewidth, on its own, \emph{does not buy finite controllability}.
Strikingly, the counterexample is tiny: the ruleset $\rs$ of \cref{sec:counterexample} consists of just three rules over a binary signature, it is pieceful, and hence \gbts and \bts, yet it is \emph{not} \fc.
A single ruleset thus settles the question negatively not only for \bts, but already for its far more restrictive subclasses \gbts and pieceful.

The counterexample also shows where finite controllability breaks: rule $\rho_3$ propagates information over unbounded distances, which is harmless in the chase but fatal in every finite model. A natural remedy is to restrict such long-distance propagation, for instance by requiring the ruleset to be \emph{local}: there is $\ell \in \nats$ such that for every database $\db$, every atom of $\chasedr$ belongs to $\chase{\db', \rs}$ for some $\db' \subseteq \db$ with $|\db'| \leq \ell$~\cite{Ostropolski2022}.
Whether locality itslef implies \fc is open, and so is the relationships between local \bts rulesets and \fc.


%

\section{AI Use Disclaimer}
LLMs have been used for drafting and editorial purposes. Authors obviously retain full responsability of the submitted content.

\bibliographystyle{kr}
\bibliography{_bibliography}

\appendix
\section{Full Proofs}\label{sec:full-proofs}

\subsection{The Proof of \cref{obs:chase-def}}\label{apx:chase-def}

\begin{proof}
    By induction on $i$. For $i = 0$ the instance in \cref{obs:chase-def} equals $\db_0 = \stepo{0}$. Assume the claim holds for $i$ and let $I = \stepo{i}$. The $\E$-atoms of $I$ are $\E(t_j, t_{j+1})$ and the $\same$-atoms are $\same(t_j, t_j)$, for $j \leq i$. In particular, $t_0$ has no $\E$-predecessor in $I$, and each $t_j$ with $1 \leq j \leq i + 1$ has the unique $\E$-predecessor $t_{j-1}$. We consider the triggers $(\rho, h)$ for $I$:
    \begin{itemize}
        \item $\rho = \rho_1$, $h(x) = t_j$, $h(y) = t_{j+1}$ with $j \leq i$: inactive for $j < i$ as $\E(t_{j+1}, t_{j+2}) \in I$; active for $j = i$.
        \item $\rho = \rho_2$, $h(x) = t_j$, $h(y) = t_{j+1}$ with $j \leq i$: inactive for $j < i$ as $\same(t_{j+1}, t_{j+1}) \in I$; active for $j = i$.
        \item $\rho = \rho_3$: then $h(y) = h(y') = t_j$ for some $1 \leq j \leq i$, so $h(x) = h(x') = t_{j-1}$ by uniqueness of $\E$-predecessors; inactive as $\same(t_{j-1}, t_{j-1}) \in I$.
    \end{itemize}
    Hence $\stepo{i+1} = \nextStep{I} = I \cup \set{\E(t_{i+1}, t_{i+2}), \same(t_{i+1}, t_{i+1})}$, where $t_{i+2}$ is the fresh null introduced by $\rho_1$. This is the required form for $i + 1$.
\end{proof}

\end{document}